# Management of Location Based Advertisement Services using Spatial Triggers
## in Cellular Networks

M. Irfan [1], M.M. Tahir N. Baig [2], Raheel M. Hashmi [3], Furqan H. Khan[4], Khurram Shehzad, Assad Ali

Department of Electrical Engineering, COMSATS Institute of Information Technology, Islamabad, Pakistan

[3] Department of Electronics & Information, Politechnico di Milano, Italy

*Abstract*- **This paper discusses the advent of new technologies which have emerged under the area of Location Based Services (LBS). An innovative implementation and approach has been presented for design of applications which are inventive and attractive towards the user. Spatial Trigger is one of the most promising additions to the LBS technologies. This paper describes ways in which mobile advertisement services can be introduced effectively in the cellular market by bringing innovation in them through effective usage of Spatial Triggers. Hence, opening new horizons to make the consumer cellular networks, commercially, more effective and informative.**

## I. INTRODUCTION

Location Based Services (LBS) are one of the most highly sought services after Value Added Services (VAS), which are targeted to generate heavy revenues for the cellular communication industry. These services, on the other hand, are aimed to benefit the user by providing valuable information and opportunity access at the same time. Bounding the discussion, it can be derived that "Location Based Services are subsidiary options to voice and data communication which employ the consumers' locations to provide them with different kinds of information services". The LBS where first introduced in the last decade of 20th century, but are still not as popular as Value Added Services (VAS) and have yet to go a long way. As the cellular communications have progresses, the advancement of LBS has also progressed. LBS are termed to be very low-cost and efficient data services which can be beneficial for the consumers as well as the network.

Location determination technology (LDT), such as Cell ID, A-GPS, E-OTD, etc., are used to find the user's location information which usually consists of X-Y coordinates [1]. For implementation of a specific location based service, modifications are be made at either the network terminals or in the mobile station (MS) equipment. In some cases, it is needed to upgrade both the network and MS for LBS implementation; however, the updates are software based solutions and involve very low enhancement costs. Some of the main service categories for LBS include Emergency and Safety services, Information and Navigation services, Tracking and Monitoring services, and Communities and Entertainment based services.

## II. LOCATION DETERMINATION TECHNIQUES

Location Determination Techniques (LDT) are an important part of LBS. Various position determination methods used include satellite based positioning; network based positioning and local positioning methods. Each of them has its merits and de-merits but almost all of them serve their purpose which is to provide the latest information about user's location. Some of the most common positioning methods with their accuracy levels are listed below.

### A. Cell Identification or Cell Global Identity (CGI):

CGI is the most basic method of mobile positioning; it is supported by all the handsets and provides the location of the mobile station based on the location of base station it is connected with [1], [2]. CGI is most commonly used alongside timing advance, together named CGI-TA. The accuracy of this method depends on the cell size. It can provide accuracy ranging from 100m to 1100m in urban areas, while its accuracy is much lower in rural areas where the cell

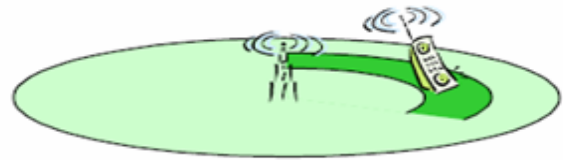

Figure 1: Cell site with Sector and Timing Advance

size is bigger [3].

### B. Enhanced Cell Global Identity (E-CGI)

In Enhanced Cell Global Identity the positioning accuracy is enhanced by making use of the power level calculated by the mobile phones together with CGI. The power level measured at the handset is used by the server to calculate distance between the base station and mobile station. As in the case of CGI, the accuracy of E-CGI also depends on cell density and can vary from 50 to 550 meters in urban areas [1],[2].

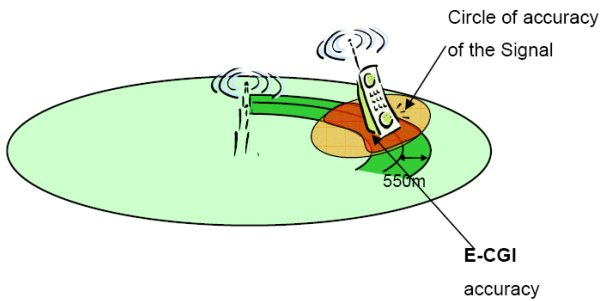

Figure 2: Enhanced Cell Global Identity

### C. Time of Arrival (TOA):

TOA determines the location of the user based on the received signal's time of arrival from three different base stations (BS) [4]. The positions of the base stations are known accurately and used to determine the position of the mobile user. The TOA method requires high synchronization between the base stations [5]. The accuracy of the location information acquired through this method ranges from 125m to 200m. The prime advantage of TOA is that it does not require extra hardware or software at MS terminal but has much greater accuracy than CGI-TA [1], [5],[6].

### D. Enhanced Observed Time Difference (E-OTD):

E-OTD is a modification of the TOA method. In E-OTD the handset measures the differences of arrival time of signals transmitted from a minimum of three synchronized base stations [5]. OTD is the time of interval that is observed by a handset between the receptions of bursts from two BS's in the cellular network. This time-measurement capability is a feature of the consumer handsets which limits this feature to only enabled and provisioned handsets to utilize the E-OTD technology. An E-OTD capable handset is equipped with special software to execute E-OTD signaling and

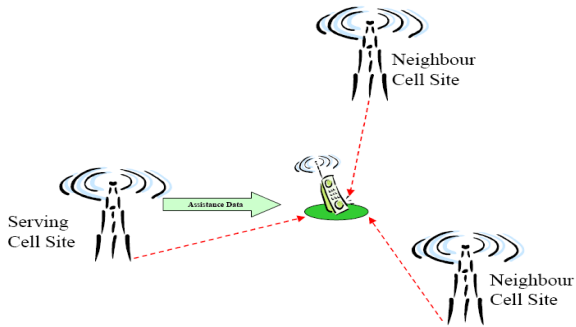

Figure 3: Enhanced Observed Time Difference

measurement methods. Time delay measurements made by the handsets are transferred via air interface to the Serving Mobile

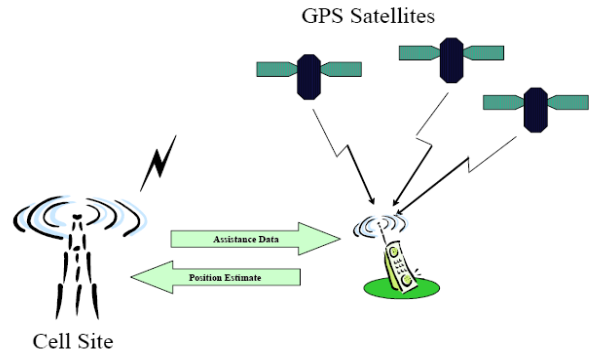

Figure 4: Assisted Global Positioning System

Location Centre (SMLC). The E-OTD method requires network modification introducing Location Measurement Units (LMU) to compensate for the case when GSM network is not very highly synchronized [2]. Accuracy of E-OTD positioning method can differ from 50 m to 150 m [3][6].

### E. Assisted Global Positioning System (A-GPS):

A-GPS is terminal based positioning technique which requires modification in both the hardware and software of the mobile handset. It is the most expensive LDT but on the other hand it is the most accurate technique with accuracies ranging from 5m to 40m [3], [5].

### III. POSITIONING REQUEST METHODS

Besides the LDT what is more important for the core network's server-end application development in LBS is the use of Location Requests. There are two major types of Location Requests discussed below.

### A. Mobile Terminated Location Request (MT-LR):

MT-LR are the requests, which arrive from outside the Public Land Mobile Network (PLMN), for the purposes like legal interception etc. These requests must come through a gate-way called GMLC (Gateway Mobile Location Centre) which verifies that the necessary agreements exist between the operator and the organization owning the external node called the LCS Client [7].

### B. Mobile Originated Location Request (MO-LR):

*MO-LR* may also come from the MS in order to support mobility applications. The procedure of MO-LR can also be used to enable a MS to request its own location to be sent to an external LCS client. The mobile initiates the location request towards the SMLC. Once location data is obtained, the MS is informed of its location and in the case where the LCS client is to be informed, the GMLC is sent a location report, which it forwards, to the LCS client [7].

As these definitions reveal, the MT-LR is more simple application and is also the one most widely used; but it does not support the high quality mobility applications as MO-LR does. The use of MO-LR can be termed as a little complex but it makes the services based on it much simple and user friendly. Moreover, the MO-LR supports the use of spatial triggers which are the most promising features of LBS since they have come into being. Figure 5 shows basic flow diagram of an MO-LR

*C. Spatial Triggers:*

Spatial Triggers are the triggers created either when a user enters or leaves a predefined geographical area, or when two MSs come relatively close to each other. Detecting such an event is the most important part of applications developed on the principal of spatial triggers. The most commonly used method to detect spatial triggers is constant database queries based on the latest location data received from the SMLC. Many companies have included the feature of spatial triggers in their GMLCs but if the system supports MO-LR, the spatial triggers can also be checked out of the GMLC. In this case the process has a slight amount of additional processing load.

The proposed approach defines the functionality of spatial triggers to ensure their best utilization to introduce location based advertisement services in commercial GSM, UMTS and other consumer cellular networks.

IV. SPATIAL TRIGGER BASED MOBILE ADVERTISEMENT

Mobile advertisements are very common during the present days. Most of these advertisements are for general purpose as they are not targeted to a single user class. The introduced approach is to develop an application which takes advantage of the user's location to send advertisements of the nearest commercial opportunities and prospective commercial outlets. Moreover, the application should also keep in view the interests of the user and the revenues of the network.

Although the developed application is for operation at the server-end yet it has three distributed parts: 1) server module, 2) advertiser handler and 3) MS module or user module. The application has to be programmed on the server equipment and is implemented as a part of GMLC. The description of these modules is given in this section in increasing order of their complexity.

*A. Advertiser Handler:*

The advertiser is the first contributor to the proposed application. To make our application attractive to the advertiser, a web interface has been designed, which provides the platform to the advertiser for management of the advertisements. Each advertiser is provided a unique identifier to login to this interface to insert the data of the prescribed, outlet or product, into the database which keeps the spatial data about all the advertisers in a certain area. This data includes advertiser's location specifications based on a pre-programmed geographical map in the application, the identifier which helps to login to the system, the service type which is to be provided and miscellaneous promotional information to be forward to the consumers. All this information can be modified by the advertiser based on the needs and desires. The selections can also be modified whenever desired using the advertiser ID.

*B. User Module:*

Each user in the customer database is requested to subscribe for this application based on individual desire and need. If a user is interested in subscription, the classification is done into a common user, GPRS user or a GPRS and GPS user. The GPRS users must also have the software package for digital mapping to earn additional benefits. Each user is also queried regarding the kind of service advertisements intended to be received. The user has the freedom to choose all, some or one of the offered advertisement classes, as per the individual aspirations. After the subscription is done, the user's location is constantly updated to the application by using MO-LR.

*C. Server Module:*

Server module is principal area for our application. It consistently maintains a database which has three major fields:

*1) User:* this section keeps the information about the user, the designated interests and the user class.

*2) Advertisement:* this section keeps the information about the advertisers, their location specifications in terms of coordinates, service types and promotional information. This information can only be modified by the advertiser in person.

*3) Info-log:* this section keeps volatile data about the user's most recent location. This data is acquired from the GMPC and is removed as soon as it is processed to remove storage overheads.

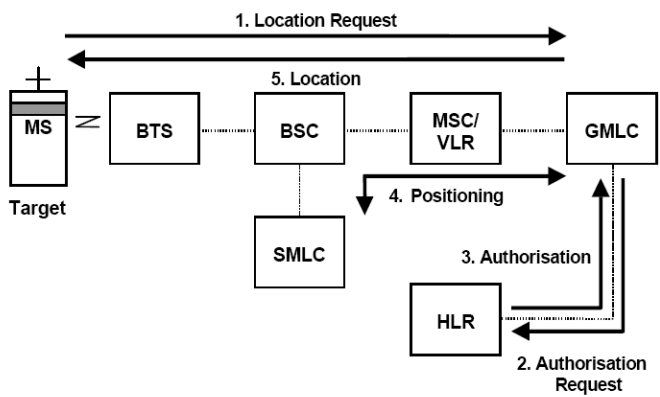

Figure 5: Mobile Originated Location Request

At present, all the service providers equipped with LBS have different features in their GMLC. Some of the GMLCs support MO-LR while some are not capable to do so. Similar is the case of spatial triggers associated with LBS. A very few GMLC's support this feature at present for example, the Ericsson's Mobile Positioning System (MPS) is one the few which not only supports MO-LR but its GMPC (Gateway Mobile Positioning Center) also checks from the spatial triggers according to pre-defined parameters. This lessens the burden on the server several orders than if the spatial triggers are created separately in a neighborhood application. Due to scarcity of such features, this research and development venture also explains the process of creating and determining spatial triggers at the application level out of the GMLC.

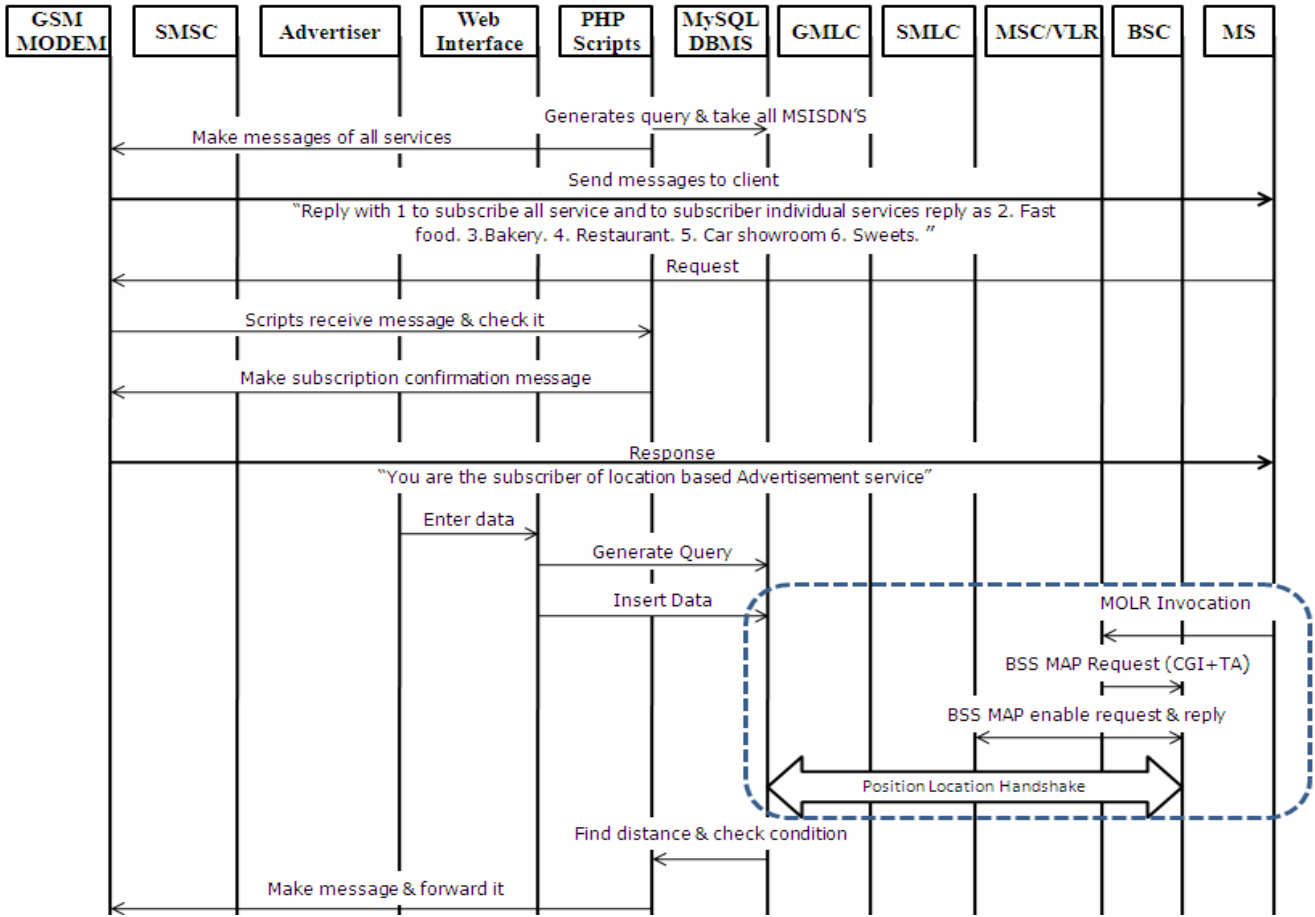

Figure 6. Message flow for operation of LBS advertisement application. The dotted region is showing MO-LR Process

The message flow in the system for operation of our application is shown in figure 1. As we start the mobile user who has subscribed for the advertisement service moves from one location to another, MO-LR is invoked which is processed by the SMLC and the location information is forwarded to the GMLC. This information contains the user's MS-ISDN, location coordinates of user and the parameters which define uncertainty in case the information has been acquired through a network based LDT. For example for a CGI-TA method it contains parameters such as inner radius, outer radius and arc width to define the area where user is located with reference to the BS.

The information forwarded by the SMLC is then forwarded to the database info-log section. This information is then picked up by the scripts which run continuously in the server. These scripts are designed and programmed using integrated JAVA and PHP support. The scripts classify the information and then check the service types against which the user has subscribed. The script then selects the location data of advertisers in that area, one at a time, calculates the distance between the user and the advertiser's coordinates and determines whether the service type matches the user's interests or not. If the distance is less than a specified limit and the service types match then it is concluded that a trigger exists and the user is to be forwarded with the designated information. It is also to be kept in notice that the spatial triggers cannot be retrieved at any level other than their mother application and hence are determined through constant query method. Once it is determined that a trigger exists between the user and the advertiser, the next step is to determine how to forward the information to the MS. As described above, we classified the user in three unique classes. The common user is forwarded the information through simple text message based on 'Flash' message format. Such a user does not have support for a high accuracy but as the service is aimed to operate in urban and densely populated areas, the uncertainty can easily be covered in approximations.

The second and third type of users can be forwarded the selected information by first determining whether their application is active. If the check results positive, the information is forwarded using the text message format with highly accurate service. The information which the user is forwarded with contains the approximate distance from the advertiser's outlet, advertisers ID and any other advertisement which the advertiser wants to attach with the message.

The scenario shown in the figure 1 has been designed and simulated using the Ericsson's Mobile Positioning System Software Development Kit (MPS-SDK) and MPC Map Tool. The map tool can be used to create route files on any given map. These route files are then loaded into the MPS-SDK whose emulator simulates them and provides the application with MO-LR based location information of the users defined in the route-file. The information is based on any available type of LDT and contains all the content types of data which SMLC forwards to the GMLC. This information is then processed as defined earlier in the section. For simulation purposes the SMSC is replaced by the NowSMS® Gateway.

## V. CONCLUSION

The spatial trigger based mobile advertisement is a unique idea which can be implemented in any environment where the GMLC supports MO-LR. This is a three tier application with all the stakeholders that is mobile user, service provider and advertisers; actively participating in the application process. As this application is applicable to all types of users, it can prove to be a good source of generating revenue for the service providers; a new and innovative platform for small business enterprise to advertise themselves; and a good, attractive, easy to use, and low cost application for the user. Moreover, the ease of integration of this application in 2G, 3G and 4G communication networks endorses the reliability and capability of this application interface.

## VI. FUTURE WORK

The applications based on the spatial triggers have a vast scope in the mobile market. With slight modifications in the database and server application this feature can also be used to introduce proximity teller services and tracking services.

The same application can also be modified to support more features in the future. The advertisements can be made more practical by allowing users to go for subscriptions offered in the advertisements by using the same service with which they are forwarded the advertisement.

ACKNOWLEDGMENT

We would like to thank the officials of Ericsson Inc. Pakistan who provided us the platforms and software support to do this R&D venture. We would also like to acknowledge the role of Mr. Riaz Hussain, Assistant Professor, CIIT Islamabad, who helped us during the course of our project.

AUTHORS PROFILE

**Muhammad Irfan**, **Mirza Muhammad Tahir Naveed Baig** and **Furqan Hameed Khan,** have done Electrical Engineering with majors in Telecommunications from Dept. of Electrical Engineering, CIIT, Islamabad in 2009. They are graduate students and are involved in research regarding the field of Value Added Services for Mobile Communications and Computer Networks.

**Raheel Maqsood Hashmi** is a graduate student at Dept. of Electronics & Information, Politecnico di Milano, Italy. He did his degree in Electrical Engineering from CIIT, Islamabad in 2009 and received Gold Medallion Award. He has research contributions in the area of Mobile Communication, Wireless Networking and Security.

**KhurramShehzad** and **Assad Ali** have done Electrical Engineering with majors in Telecommunication from Dept. of Electrical Engineering, CIIT, Islamabad in 2009. They were recommended by CIIT, EE Dept as student researchers for the CIMI (CIIT Medals for innovation) Awards 2008. They have research contributions in the area of Mobile Communication and QoS Management in Wireless Networks.